\documentclass[12pt,a4paper,english,amssymb,nofootinbib,superscriptaddress]{revtex4}
\usepackage{lmodern}
\usepackage{lmodern}
\usepackage[T1]{fontenc}
\usepackage[latin9]{inputenc}
\usepackage{color}
\usepackage{babel}
\usepackage{verbatim}
\usepackage{amsmath}
\usepackage{amssymb}
\usepackage{esint}
\usepackage[unicode=true,pdfusetitle,
 bookmarks=true,bookmarksnumbered=false,bookmarksopen=false,
 breaklinks=false,pdfborder={0 0 1},backref=false,colorlinks=true]
 {hyperref}
\hypersetup{
 linkcolor=blue,citecolor=blue}

\makeatletter

\pdfpageheight\paperheight
\pdfpagewidth\paperwidth

\@ifundefined{textcolor}{}
{%
 \definecolor{BLACK}{gray}{0}
 \definecolor{WHITE}{gray}{1}
 \definecolor{RED}{rgb}{1,0,0}
 \definecolor{GREEN}{rgb}{0,1,0}
 \definecolor{BLUE}{rgb}{0,0,1}
 \definecolor{CYAN}{cmyk}{1,0,0,0}
 \definecolor{MAGENTA}{cmyk}{0,1,0,0}
 \definecolor{YELLOW}{cmyk}{0,0,1,0}
}

\usepackage{babel}
\usepackage{babel}
\usepackage{babel}
\usepackage{babel}
\usepackage{babel}
\usepackage{babel}
\@ifundefined{definecolor}{color}{}
\usepackage{babel}

\usepackage{latexsym}\usepackage{bm}

\makeatother

\begin{document}

\title{Classical Double Copy: Kerr-Schild-Kundt metrics
from Yang-Mills Theory
}

\author{Metin G{\"u}rses}
\email{gurses@fen.bilkent.edu.tr}

\selectlanguage{english}%

\affiliation{{\small{}Department of Mathematics, Faculty of Sciences}\\
 {\small{}Bilkent University, 06800 Ankara, Turkey}}

\author{Bayram Tekin}
\email{btekin@metu.edu.tr}

\selectlanguage{english}%

\affiliation{Department of Physics,\\
 Middle East Technical University, 06800 Ankara, Turkey}
\begin{abstract}
\noindent The classical double copy idea relates some solutions of Einstein's theory with those of gauge and scalar field theories. We study the Kerr-Schild-Kundt (KSK) class of metrics
in $d$-dimensions in the context of possible new examples of this idea. We first show that it is possible to solve the Einstein-Yang-Mills system exactly using the solutions of a Klein-Gordon type scalar equation when the metric is the $pp$-wave metric which is the simplest member of the KSK class. In the more general KSK class, the solutions of a scalar equation also solve the Yang-Mills, Maxwell and Einstein-Yang-Mills-Maxwell equations exactly albeit with a null fluid source. Hence in the general KSK class, the  double copy correspondence is not as clean-cut as in the case of the $pp$-wave. In our treatment all the gauge fields couple to dynamical gravity, and are not treated as test fields. We also briefly study G\"{o}del type metrics along the same lines.
\end{abstract}
\maketitle

\section{Introduction}
Constructing solutions of Einstein field equations, with a source or in a vacuum, is so difficult that
anytime a new method is suggested, one should embrace it with enthusiasm.
The recent "classical double copy" correspondence \cite{mon} is such a new idea
which we shall pursue here for some {\it exact } gravity waves in the hope of extending the earlier examples \cite{mar}.  The basic essence of the classical double copy method is this: one can find
some classical solutions of general relativity  from the classical solutions of Yang-Mills or Maxwell field equations or even from those of a simpler scalar field equation. This construction, gravity being a double copy of the YM theory- which
itself is a single copy-and a scalar field (usually a bi-adjoint real
scalar field)-which is the zeroth copy-is an extension of a powerful idea
and observation  that goes beyond the classical level: the scattering
amplitudes in general relativity and those of {\it two copies} of Yang-Mills theories are related. This is known as the Bern-Carrasco-Johansson (BCJ)  double copy \cite{bern} and works perturbatively granted that the color and kinematic factors are identified accordingly. For more details and the references, see \cite{godoy,Lee}.

The classical double copy correspondence has been mostly studied in the Kerr-Schild class of metrics. This class has remarkable properties and includes a large number of physical metrics, including the Kerr black hole. In this work, to extend the set of examples and to understand the possible limitations to the classical double copy correspondence, we study a  class of spacetime, the so called Kerr-Schild-Kundt (KSK) class, which turned out to be universal, in the sense that KSK-metrics solve all metric-based theories  \cite{gur1,gur2,Gurses-PRL,gur3,smooth,gur4}. The classical double copy arguments make use of the metric in the Kerr-Schild form
\begin{equation}\label{ksk}
g_{\mu \nu}=\bar{g}_{\mu \nu}+2 V \ell_{\mu} \ell_{\nu},
\end{equation}
where $\bar{g}_{\mu \nu}$ is the background (or the the seed) metric, $\ell_{\mu}$ is a null vector with respect to both metrics and $V$, at this stage, is an arbitrary function. The fact that $\ell_{\mu}$ is null is a crucial point in what follows. In fact to see this explicitly, for the moment let us assume that it is {\it not} null, then the inverse metric reads
\begin{equation}
g^{\mu \nu}=\bar{g}^{\mu \nu}-\frac{2 V}{ 1 + 2  V \,\ell^2}  \ell^{\mu} \ell^{\nu},
\end{equation}
where $\ell^2 \equiv \bar{g}^{\mu \nu} \ell_{\mu} \ell_{\nu}$. It is clear that only for the null case, the inverse metric is linear in the metric profile function $V$, a fact  that dramatically simplifies all the ensuing discussion. [Note that in the last part of this work, we briefly consider the metrics that are defined with a non-null vector field.]

In the works on classical double copy,  one usually encounters the following construction: the seed metric is taken to be flat, namely $g_{\mu \nu}=\eta_{\mu \nu}+2 V \ell_{\mu} \ell_{\nu}$, the Maxwell and Yang-Mills fields are taken as $A_{\mu}=V\, \ell_{\mu}$ and the Yang-Mills field  ${\cal{A}}^{a}_{\mu}=c^{a} V \ell_{\mu}$ where $c^{a}$'s are constants. In this case, one can only treat the Maxwell and the Yang-Mills as test fields. If the metric satisfies the vacuum field equations ($G_{\mu \nu}=0$) then the spacetime becomes stationary {\it i.e.} $\partial_0 V=0$ \cite{mon,mar} and the metric function satisfies the Laplace's equation $\nabla^2 V=0$. Any solution of this equation also solves the Maxwell and Yang-Mills equations identically. This construction was extended to the maximally symmetric non-flat backgrounds in \cite{mar}. If one relaxes the stationarity  assumption, {\it i.e.} $\partial_{0} V \ne 0$,  but instead imposes the constraint $\ell^{\mu} \partial_\mu=0$, one obtains a nice exact result which supports the classical double copy approach.

The lay-out of this work is as follows: we first start with the $pp$-waves and give a solution of the coupled Einstein-Yang-Mills-Maxwell system that is in the double copy spirit. We then discuss a possible extension to the general KSK class and show that a null fluid is needed for that case for the correspondence to work. Our results are summarized in two theorems.  In the conclusion  and further discussions part we also discuss a possible extension of these ideas to the G\"{o}del-type metrics with non-null vector fields. The motivation is to extend the double copy correspondence possibly to the massive gauge field case.

\section{KSK Metrics and double copy}

Our first main result is on the exact solutions of the Einstein-Yang-Mills-Maxwell field equations where the spacetime is the $d$-dimensional $pp$-wave geometry. We first state our main results for the {\it pp} waves as a theorem and provide the proof later as a subclass of the KSK case.

Setting all relevant coupling constants to unity, the coupled Einstein, Maxwell and Yang-Mills equations  are
 \begin{eqnarray}
 && G _{\mu \nu}=\gamma_{ab}\, {\cal{F}}^{a\, \alpha}\,_{\mu}\, {\cal{F}}^{b}_{\alpha \nu}-\frac{1}{4}{\cal{F}}^2 g_{\mu \nu}+
\sum_{k=1}^{N}\,\left( F^{k\,\alpha}\,_{\mu}\, F^{k}_{\alpha \nu}-\frac{1}{4}{F^{k}}^2 g_{\mu \nu}\right) \label{ein0},\nonumber \\
&& \nabla_\mu F^{k\,\,\mu \nu} =0, \hskip 1 cm {(D_\mu {\cal{F}}^{\mu \nu})}^a =0,
\label{system1}
 \end{eqnarray}
where the gauge-covariant derivative is $D_\mu \equiv I \nabla_\mu - i T^a {\cal {A}}_\mu^a$  with the generators satisfying $[T^a, T^b] = i f^{abc} T^c$ and the inner product is taken as tr $(T^a T^b) = \frac{1}{2} \gamma^{a b}$.\footnote{We do not specify the underlying Lie algebra of the non-abelian theory, but it can be taken to be any non-abelian Lie algebra.}. We assume there are  $N$ number of Maxwell's field $F^{k}_{\mu \nu}, k=1,2,\cdots, N$.

\vspace{0.5cm}
 \noindent
 Let us take the spacetime be the $d$-dimensional {\it pp}-wave geometry with the metric given in the Kerr-Schild form as $g_{\mu \nu}=\eta_{\mu \nu}+2 V \ell_{\mu} \ell_{\nu}$ where $\ell_{\mu}$ is a covariantly constant null vector, and let $A^{k}_{\mu}=\phi^{k} \, \ell_{\mu}, k=1,2, \cdots, N$ be  abelian and ${\cal{A}}^{a}_{\mu}=\Phi^{a} \, \ell_{\mu}$ be
non-abelian vector potentials satisfying the properties
 \begin{equation}
 \ell^{\mu}\, \partial_\mu \phi^{k}=\ell^{\mu}\, \partial_\mu \Phi^{a}=0.
 \end{equation}
Then, one can show that the Einstein tensor reduces to
\begin{eqnarray}
G_{\mu \nu}= -\ell_{\mu}\, \ell_{\nu}{\bar \square}\,V,
\end{eqnarray}
while the Maxwell and Yang-Mills field equations reduce to
\begin{equation}
\bar{\square} \phi^{k}=\bar{\square} \Phi^{a}=0,
\label{scalar_eqns}
\end{equation}
where $\bar{\square} \equiv \eta^{\mu \nu}\partial_\mu \partial_\nu$. We can now state the first theorem.

{\bf Theorem 1:}  {\it Under the assumptions made above, the field equations (\ref{ein0}) reduce to
\begin{equation}
{\bar \square} V=-\bar{g}^{\mu \nu}\, \left[\gamma_{ab} \partial_{\mu} \Phi^{a}\, \partial_{\nu} \Phi^{b}+\sum_{k=1}^{N}\, \partial_{\mu} \phi^{k}\, \partial_{\nu} \phi^{k} \right], \label{eqn1}
\end{equation}
whose most general solution is
 \begin{equation}\label{denk0}
 V=V_{0}+c_{k}\, \phi^{k}+ \beta_{a} \Phi_{a}-\frac{1}{2}\,\gamma_{ab}\,\Phi^{a}\, \Phi^{b}-\frac{1}{2} \sum_{k=1}^{N}\, \phi^{k}\, \phi^{k},
\end{equation}
where  $V_0$ is the vacuum solution satisfying ${\bar{\square} } V_{0}=0$; and $c_{k}$, $\beta^a$ are arbitrary constants and $\phi^{k}$ and $\Phi_{a}$ satisfy (\ref{scalar_eqns}).}

Given any solution of (\ref{scalar_eqns}), and there are many, one can find the corresponding metric via the profile function (\ref{denk0}). Observe that if one further takes  $\Phi^{a}=t^a \phi$ and $\phi^k=p^k \phi$ where $t^a$ and $p^k$ are constants, then one only needs to solve a single scalar equation $\bar{\Box} \phi=0$ for $\phi$. Let us note that  this solution generalizes the solutions of \cite{mon} and \cite{mar} where $V_{0}=0$ and the gauge fields are treated as test fields that do not change the background geometry, but here we have given the solution of the full coupled system. In the rest of the paper, we do not explicitly consider the Maxwell fields, but  embed them in the Yang-Mills fields by enlarging the gauge group.  For this purpose, we let the Maxwell fields vanish without losing any generality.

Our next task is to try to generalize this result for the general KSK class, which has been studied in some detail recently in \cite{gur1}-\cite{gur4}. In generalized Kerr-Schild coordinates, the metric is taken to be\footnote{We use the vector $\lambda_\mu$ for this case instead of the previous $\ell_\mu$ as we shall reserve $\ell$ for the AdS radius.}
\begin{equation}
g_{\mu\nu}=\bar{g}_{\mu\nu}+2V\lambda_{\mu}\lambda_{\nu},\label{eq:AdS-waveKS}
\end{equation}
where the seed $\bar{g}_{\mu\nu}$ metrics are maximally symmetric. One
can show that the following relations hold for the metrics belonging
to the KSK class:
\begin{equation}
\lambda^{\mu}\lambda_{\mu}=0,\qquad\nabla_{\mu}\lambda_{\nu}\equiv\xi_{(\mu}\lambda_{\nu)},\qquad\xi_{\mu}\lambda^{\mu}=0,\qquad\lambda^{\mu}\partial_{\mu}V=0.\label{eq:AdS-wave_prop}
\end{equation}
The first property is the usual nullity condition of the vector, while the second and the third ones guarantee that the $\lambda$ vector is geodesic $\lambda^{\mu}\nabla_\mu \lambda_\nu=0$. These three conditions define the KSK class of metrics. The last property is required for further simplifications, such as the linear dependence of the mixed Einstein tensor on $V$. For more on this  point in the context of Kerr-Schild metrics, see \cite{Gurses_Gursey} for the flat seed and \cite{Dereli} for the generalized cases. For this class of metrics,  the traceless-Ricci tensor, $S_{\mu\nu}\equiv R_{\mu\nu}-\frac{R}{d}g_{\mu\nu}$,
can be computed to yield
\begin{equation}
S_{\mu\nu}=\rho \lambda_{\mu}\lambda_{\nu},~
\end{equation}
where the scalar function $\rho$ is found to be linear in $V$ which reads explicitly
\begin{equation}
\rho=-\left(\bar \square+2\xi^{\mu}\partial_{\mu}+\frac{1}{2}\xi^{\mu}\xi_{\mu}-\frac{2\left(d-2\right)}{\ell^{2}}\right)V\equiv-{\cal Q}_{1} V.\label{rho}
\end{equation}
We defined the operator ${\cal Q}_1$ in the second equality. The Weyl tensor, $C_{\mu\alpha\nu\beta}$, can be found to be \cite{Gurses-PRL},
\begin{equation}
C_{\mu\alpha\nu\beta}=4\lambda_{[\mu}\Omega_{\alpha][\beta}\lambda_{\nu]},\label{eq:Weyl_KSK}
\end{equation}
where the square brackets denote anti-symmetrization with a $1/2$
factor and the symmetric tensor $\Omega_{\alpha\beta}$
is a rather non-trivial object, but it is still linear in $V$ and can be compactly written as
\begin{equation}
\Omega_{\alpha\beta}\equiv-\left[\nabla_{\alpha}\partial_{\beta}+\xi_{(\alpha}\partial_{\beta)}+\frac{1}{2}\xi_{\alpha}\xi_{\beta}-\frac{1}{d-2}g_{\alpha\beta}\left({\cal {Q}}_1+\frac{2\left(d-2\right)}{\ell^{2}}\right)\right]V.
\end{equation}
Form the Weyl tensor and the traceless Ricci tensor given here, one can compute the  needed curvature invariants for these metrics.
As two examples of the KSK class, let us give the AdS plane and the AdS spherical wave metrics which read:

\vspace{0.5cm}
\noindent
{\bf AdS plane wave metrics}
\begin{equation}
ds^2=\frac{\ell^2}{z^2}\, \left(2 du dv+\sum_{m=1}^{d-3} (dx^{m})^2+dz^2 \right)+2V(u,x^{m}, z) du^2,
\end{equation}
where $z = x^{d-1}$ , $\lambda_{\mu}\,dx^{\mu}=du$ and $\xi_{\mu} dx^{\mu}=\frac{2}{z}\, dz$.

\vspace{0.5cm}
\noindent
{\bf AdS spherical wave metrics:}
\begin{equation}
ds^2=\frac{\ell^2}{z^2}\, \left(-dt^2+\sum_{m=1}^{d-1} (dx^{m})^2\right)+2V(t,x^{m}, z) du^2,
\end{equation}
where
\begin{equation}
\lambda_{\mu}\,dx^{\mu}=dt+\frac{1}{r}\, \vec{x} \cdot d\vec{x},  \hskip 0.5 cm \xi_{\mu}dx^{\mu}=-\frac{1}{r} \lambda_{\mu}\,dx^{\mu}+\frac{2}{r} dt+\frac{2}{z}.
\end{equation}
Here $r^2=\sum_{m=1}^{d-1}\, (x^{m})^2$. The function $V$ satisfies the constraint $\lambda^{\mu}\, \partial_\mu V=0$.
In \cite{Gurses-PRL}, we showed that the AdS-plane wave and the
{\it pp}-wave metrics, and more generally all KSK
metrics, are universal in the sense that they solve all metric-based gravity equations with only slight changes in the parameters, such as the cosmological constant. Different seed metrics ($\bar{g}_{\mu\nu}$) lead to different spacetimes: it is the flat Minkowski metric for the {\it pp}-waves, it is the AdS spacetime for the AdS-plane and the AdS-spherical waves,
and it is the de Sitter spacetime for the dS-hyperbolic wave. After this brief recap of the KSK metrics, we can state our second  theorem.

\vspace{0.5cm}
 \noindent
 {\bf Theorem 2} {\it Let the spacetime be the $d$-dimensional KSK geometry with the metric $g_{\mu \nu}=\bar{g}_{\mu \nu}+2 V \lambda_{\mu} \lambda_{\nu}$ and let  ${\cal{A}}^{a}_{\mu}=\Phi^{a} \, \lambda_{\mu}$ be a non-abelian vector potential, satisfying the property
 \begin{equation}\label{const}
\lambda^{\mu}\, \partial_\mu \Phi^{a}=0.
 \end{equation}
 Then the Einstein Maxwell, Yang-Mills, null dust field equations with a cosmological constant
 \begin{eqnarray}\label{ein1}
 && G _{\mu \nu}=\gamma_{ab}\, {\cal{F}}^{a\, \alpha}\,_{\mu}\, {\cal{F}}^{b}_{\alpha \nu}-\frac{1}{4}{\cal{F}}^2 g_{\mu \nu}-\Lambda g_{\mu \nu}+\varepsilon u_{\mu} u_{\nu}, \nonumber \\
&& {(D_\mu {\cal{F}}^{\mu \nu})}^a =0, \hskip 1 cm \nabla^\mu (\varepsilon u_{\mu} u_{\nu}) =0,
\label{system2}
\end{eqnarray}
have the solution
 \begin{eqnarray}
 &&V= \beta_{a} \Phi_{a}-\frac{1}{2}\,\gamma_{ab}\,\Phi^{a}\, \Phi^{b}, \label{denk1} \nonumber \\
&&\varepsilon= \left(\xi^{\mu} \partial_{\mu}+\frac{1}{2} \xi^{\mu}\, \xi_{\mu}-\frac{2(d-2)}{\ell^2} \right) \left( \frac{1}{2}\gamma_{ab}\, \Phi^{a}\, \Phi^{b}+\beta_{a}\, \Phi^{a} \right), \label{denk2} \nonumber \\
&&\Lambda=-\frac{(d-1)(d-2)}{2 \ell^2}, \nonumber \\
&&u_{\mu}=\lambda_{\mu}.
\end{eqnarray}
}
The Einstein tensor takes the form
\begin{equation}
G_{\mu \nu}= - \lambda_{\mu}\, \lambda_{\nu}{\cal Q}_{1} \,V\,\,+\frac{(d-1)(d-2)}{2 \ell^2} g_{\mu \nu},
\end{equation}
while the Yang-Mills  equation reduces to
\begin{equation}
{\cal{Q}}_{2} \Phi^{a}=0,
\end{equation}
where ${\cal Q}_{2} \equiv \bar{\square}+\xi^{\mu}\, \partial_{\mu}$.

The proof of this theorem is as follows: the field strength of the Yang-Mills fields can be computed to be
\begin{eqnarray}
&&{\cal{F}}^{a}_{\mu \nu}=\partial_\mu \Phi^{a} \lambda_{\nu}-\partial_\nu \Phi^{a} \lambda_{\mu},
\end{eqnarray}
whose nonlinear part vanishes. Using the assumption (\ref{const}), one finds 
\begin{eqnarray}
&&\nabla^{\mu}\, {\cal{F}}^{a}_{\mu \nu}=\left (\bar{\square} +\xi^{\alpha}\, \partial_{\alpha}\right)\, \Phi^{a} \, \lambda_{\nu}=0,
\end{eqnarray}
which then leads to
\begin{equation}
{\cal Q}_{2} \Phi^{a}=0.
\end{equation}
The energy-momentum tensor of the gauge field becomes
\begin{eqnarray}
&&T^{\mbox{YM}}_{\mu \nu}={\bar g}^{\alpha \beta}\, \gamma_{ab}\, \partial_\alpha\Phi^{a}\, \partial_\beta \Phi^{b}\lambda_{\mu} \lambda_{\nu}.
\end{eqnarray}
Then the field equations (\ref{ein1}) reduce to
\begin{eqnarray}
-{\cal Q}_{1} V&=& {\bar g}^{\alpha \beta}\, \gamma_{ab}\,\partial_\alpha \Phi^{a}\, \partial_\beta \Phi^{b}+\varepsilon, \nonumber \\
\Lambda&=-&\frac{(d-1)(d-2)}{2 \ell^2}.
\end{eqnarray}
Moreover, one can show that
\begin{equation}
{\cal Q}_{2}\, \left(\frac{1}{2}\, \gamma_{ab}\, \Phi^{a}\, \Phi^{b} \right)= {\bar g}^{\alpha \beta}\, \gamma_{ab}\, \partial_\alpha \Phi^{a} \, \partial_\beta \Phi^{b}.
\end{equation}
Hence one obtains
\begin{equation}
-{\cal Q}_{2} \left(V+\frac{1}{2}\, \gamma_{ab}\, \Phi^{a}\, \Phi^{b} \right)= \left(\xi^{\mu} \partial_{\mu}+\frac{1}{2} \xi^{\mu}\, \xi_{\mu}-\frac{2(d-2)}{\ell^2} \right)\, V+\varepsilon  \label{denk03}
\end{equation}
Assuming that both sides of (\ref{denk03}) vanish, then one obtains (\ref{denk1}).
Note that the solution of a single equation ${\cal Q}_{2} \Phi^{a}=0$ solves all the Einstein-Yang-Mills and null dust field equations identically. Ignoring the quadratic terms in $V$ we obtain solutions of the Einstein field equations where the Yang-Mills field is a test field. Vanishing of the vector $\xi_{\mu}$ means that the vector $\lambda_{\mu}$ becomes a covariantly constant vector field. In a spacetime with such a vector field, the cosmological constant vanishes identically and the metric reduces to the $pp$-wave metric.  Then for vanishing $\xi$, the null dust also vanishes, {\it i.e.}, $\varepsilon=0$,  then Theorem 2 reduces to Theorem 1 and hence the proof of Theorem 1 also follows.

For this brief part, let us assume that we have a single Maxwell field and a non-abelian gauge field. Then, in Theorem 1, it is be possible to introduce coupled equations between $\phi$ and $\Phi^{a}$. Let the field equations be
\begin{eqnarray}
&&D^{\mu}\, {\cal{F}}^{a}_{\mu \nu}={\cal{J}}^{a}_{\nu},  \nonumber \\
&&\nabla^{\mu}\, F_{\mu \nu}=j_{\nu},
\end{eqnarray}
Then the covariant conservation yields
\begin{equation}
\nabla^{\mu}\, G _{\mu \nu}=\gamma_{ab}\, {\cal{J}}^{a\, \alpha}\,\, {\cal{F}}^{b}_{\alpha \nu}+j^{\alpha}\,\, F_{\alpha \nu}=0.
\end{equation}
The right-hand side vanishes identically since ${\cal{J}}^{a}_{\nu}= \lambda_{\nu} \bar{\square} \Phi^{a}$ and $j_{\nu}=  \lambda_{\nu} \bar{\square} \phi$. Hence the Einstein equations in (\ref{ein0}) reduce to
\begin{equation}
-\bar{\square} V=\bar{g}^{\alpha \beta} \gamma_{ab}\, \partial_\alpha \Phi^{a}\, \partial_\beta \Phi^{b}+\bar{g}^{\alpha \beta}\, \partial_\alpha\phi \, \partial_\beta\phi,
\end{equation}
which is equivalent to the following
\begin{equation}
-\bar{\square} \left(V+\frac{1}{2} \gamma_{ab}\, \Phi^{a}\, \Phi^{b}+\frac{1}{2} \phi^2 \right)=-\gamma_{ab}\Phi^{a}\, \bar{ \square} \Phi^{b}-\phi\, \bar{\square} \phi.
\end{equation}
Hence we can let
\begin{equation}
V=c \phi+\beta_{a} \Phi^{a}-\frac{1}{2} \gamma_{ab}\, \Phi^{a}\, \Phi^{b}-\frac{1}{2} \phi^2,
\end{equation}
and
\begin{eqnarray}
&&\bar{\square} \Phi^{a}= \phi\, f^{a}(\phi, \Phi^{a}, \partial \phi, \partial \Phi^{a}), \nonumber \\
&&\bar{\square} \phi= -\gamma_{ab}\, \Phi^{b}\,f^{a}(\phi, \Phi^{a}, \partial \phi, \partial \Phi^{a}),
\end{eqnarray}
where $f^{a}(\phi, \Phi^{a}, \partial \phi, \partial \Phi^{a})$ is an arbitrary function of its arguments.  Hence we obtain a coupled system of nonlinear equations for $\phi$ and $\Phi^{a}$. We get a rather simple example by letting $\Phi^{a}=c^{a} \psi$ and $f^{a}=\kappa\, c^{a}$ then
\begin{equation}
\bar{\square} \phi=-\kappa c^2 \psi,~~~~\bar{\square} \psi=-\kappa \phi,
\end{equation}
where $c^2=c^{a}\,c^{b}\, \gamma_{ab}$. These equations can decoupled as
\begin{equation}
\bar{\square}^2 \phi=-m^2 \phi,~~~~\bar{\square}^2 \psi=-m^2 \psi,
\end{equation}
where $m^2=\kappa^2\, c^2$. Similar extension can be made in Theorem 2 as well.

\section{Conclusions and Further Discussions}

We have studied the $pp$-wave and Kerr-Schild-Kundt geometries as examples of the classical double copy correspondence in the coupled Einstein-Yang-Mills system.  For the $pp$-wave case, the metric profile function ($V$) is given as a quadratic and linear function of the scalar fields defining the Yang-Mills fields as (taking $V_{0}=0$)
\begin{equation}
V= \beta_{a} \Phi_{a}-\frac{1}{2}\,\gamma_{ab}\,\Phi^{a}\, \Phi^{b},
\end{equation}
which nicely fits in the double copy notion as gravity is basically ``square of the gauge theory".  In the general KSK case, for the double copy correspondence to work, we have shown that one also needs a null dust, a fact which somewhat complicates the correspondence. As a further extension, one might wonder how far one can go if the condition on the nullity of the Kerr-Schild vector field is relaxed. For this purpose, below is a a brief account of an attempt in such metrics.

Let $(g_{\mu \nu}, M)$ be a $d$-dimensional spacetime geometry with the metric
\begin{equation}
g_{\mu \nu}=h_{\mu \nu}-u_{\mu}\,u_{\nu},
\end{equation}
where $u_{\mu}$ is a unit  {\it timelike} vector field and $h_{\mu \nu}$ is a degenerate matrix of rank $d-1$. We let $u^{\mu}=-\frac{1}{u_{0}} \delta^{\mu}_{0}$ and $u^{\mu}\, h_{\mu \nu}=0$. The determinant of the metric is $g=-u_{0}^2$. We call such a spacetime metric as a "G{\" o}del-type metric" \cite{gs1,gs2}. Here, for a simple construction, we will assume that $u_{0}$ is a nonzero constant and $u_{\mu}$ is a Killing vector field.
We assume also that $\partial_{0}\, u_{\alpha}=0$. With these information, one can find the
field strength
\begin{eqnarray}
f_{\mu \nu} \equiv\nabla_{\mu}\,u_{\nu}-\nabla_{\nu}\, u_{\mu}=2 \nabla_{\mu}\, u_{\nu}, \hskip 0.5 cm f^{2} \equiv f_{\alpha \beta}\,f^{\alpha \beta}, .
\end{eqnarray}
and the Einstein equations as
\begin{equation}
G_{\mu \nu}=\frac{1}{2} T_{\mu \nu}+\frac{1}{4} f^2 u_{\mu}\, u_{\nu},
\end{equation}
where $T_{\mu \nu}$ is the Maxwell energy momentum tensor. We have constructed a metric which satisfies  the Einstein Maxwell-dust field equations identically provided that the vector field $u_{\mu}$ satisfies  the source-free Maxwell equation
\begin{equation}
\partial_{\alpha}\, f^{\alpha}\,_{\mu}=0.
\end{equation}
Observe that the partial derivative appears in this expression, but, under the assumptions made so far, one can show that the last equation is equivalent to the following equation
\begin{eqnarray}
&&\nabla_{\alpha}\, f^{\alpha}\,_{\mu}=\frac{1}{2} f^2\, u_{\mu},
\end{eqnarray}
or
\begin{equation}
(\square\, -\frac{1}{4} f^2 )\, u_{\mu}=0.
\end{equation}
A simpler version of this equation is $\partial_{i} f_{ij}=0$. All the above equations on the vector $u_{\alpha}$ can be simplified further. Since $u_{0}$ is assumed to be constant then $\vec{u}$ satisfies the linear equation
\begin{equation}
\nabla^2 \vec{u}-\vec{\nabla}\, (\vec{\nabla} \cdot \vec{u})=0
\end{equation}

Hence any solution of the last equation  also solves  Einstein Yang-Mills field equations identically where $A^{a}_{\mu}= c^{a}\,u_{\mu}$ and $g_{\mu \nu}=h_{\mu \nu}-u_{\mu} u_{\nu}$. When $u_{0}$ is not a constant then the metric can be extended further to a scalar (dilaton) field. G{\" o}del-type metrics can be used in solving the Einsten-Yang-Mills dilaton 3- form field equations \cite{gs2} from a single vector equation.
These metrics deserve a separate discussion which we shall give elsewhere.

\end{document}